\documentclass[prl,twocolumn,superscriptaddress,showpacs]{revtex4}

\usepackage{epsfig,amsmath,latexsym}

\begin{document}
\newcommand{\Go}{\ensuremath{\textrm G_{0}}}
\newcommand{\brho}{\mbox{\boldmath$\rho$}}

\title{Origin Of Current-Induced Forces In An Atomic Gold Wire: A First Principles Study}
\preprint{1}
\author{Mads Brandbyge}
\email[Corresponding author: ]{mbr@mic.dtu.dk} 
\affiliation{Mikroelektronik Centret
(MIC), Technical University of Denmark, Bldg.~345E, DK-2800 Lyngby,
Denmark}
\author{Kurt Stokbro}
\affiliation{Mikroelektronik Centret
(MIC), Technical University of Denmark, Bldg.~345E, DK-2800 Lyngby,
Denmark}
\author{Jeremy Taylor}
\affiliation{Mikroelektronik Centret
(MIC), Technical University of Denmark, Bldg.~345E, DK-2800 Lyngby,
Denmark}
\author{Jos\'e-Luis Mozos}
\affiliation{Institut de Ciencia de Materials de Barcelona, 
CSIC Campus de la U.A.B., 08193 Bellaterra, Spain}
\author{Pablo Ordej\'on}
\affiliation{Institut de Ciencia de Materials de Barcelona, 
CSIC Campus de la U.A.B., 08193 Bellaterra, Spain}
\date{\today}
\begin{abstract}
We address the microscopic origin of the current-induced forces by analyzing
results of first principles density functional 
calculations of atomic gold wires connected to two gold electrodes
with different electrochemical potentials. We find that current
induced forces are closely related to the chemical bonding, and
arise from the rearrangement of bond charge due to the current
flow. We explain the current induced bond weakening/strengthening by 
introducing bond charges decomposed into electrode
components.
\end{abstract}
% 
%73.40.Jn,  metal-to-metal contacts
%66.30.Qa,  electromigration
% 87.15.By  Structure and bonding
%
\maketitle
%%% Review, intro,

The failure of interconnects in integrated circuits is often due to
directionally biased diffusion of atoms caused by the presence of an
electric current (electromigration)\cite{Sorbello98}.  In recent years
the prospect of electronic devices operating essentially on the
atomic/molecular scale has gained significant interest\cite{JoGiAv00},
and electromigration and current-induced conformational changes are
important issues when downscaling electronic components to these
sizes\cite{VePaLa02}. A single atom wide gold contact is among the
simplest atomic scale conductors and therefore ideal for fundamental
studies of the current-induced forces. It is remarkable that these
systems can sustain voltages up to several volts before
breaking\cite{YaBoBr98,HaNiBr00,NiBrHa02}.

%%% This paper: 
Our aim here is to uncover the microscopic origin of the
current-induced bond weakening/strengthening in an atomic gold wire.
We analyze state-of-the-art first principles calculations
of the current-induced forces while other investigations have focused
on the theoretical foundation of these forces\cite{ToHoSu00}. 
Traditionally the microscopic origin of the current induced force on an atom is related
to electrostatic forces and the momentum transfer by the electron
flow, the so called electron wind force\cite{Sorbello98}. We find that
the current-induced forces are rather linked to the redistribution of
bond charge or overlap population in the system. By extending the
concept of overlap population to a nonequilibrium system with two
electrochemical potentials, we can rationalize the first principles
results in terms of a simple two orbital interaction model.

%% review DFT + 
%%% intro to calculation
We have performed our calculations within density functional theory
(DFT) \cite{method,SoArGa02,BrMoOr02}.  We employ the {\sc TranSIESTA}
program\cite{BrMoOr02} which allows a full atomistic description of
the scattering region and the electrodes. DFT has previously been used
to describe the current-induced forces acting on single
atoms\cite{La92,KoHiTs96}, molecules\cite{VePaLa02}, and an atom
adsorbed on an nanotube\cite{MiYaHa01}. Recently, the current-induced
forces in atomic wires, and the resulting ``embrittlement'', was
addressed by more approximate theory\cite{ToHoSu01}.

\begin{figure}[!h]
\epsfig{figure=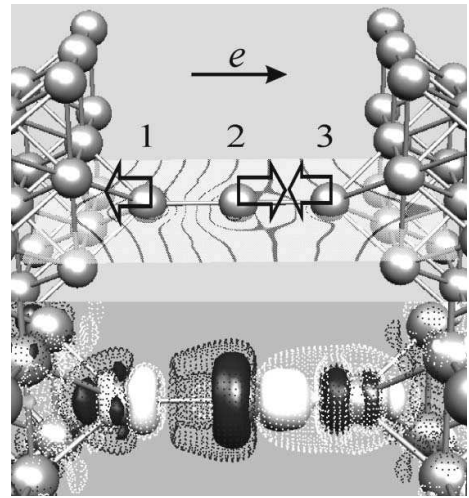, width=0.7\columnwidth, angle=0}
\caption{ (a) Direction of forces and voltage drop for the gold wire
  connecting (100) electrodes at 1V bias.  (b) Isodensity surfaces for
  the {\em change} in density from 0V to 1V. Dark is deficit and white
  is extra electron density.  The solid (dotted) surface correspond to
  $\pm 5\cdot 10^{-4} e/\mbox{\AA}^3$ ($\pm 2\cdot 10^{-4}
  e/\mbox{\AA}^3$).  }
\label{fig:1}
\end{figure}

% Setup
We consider a simple symmetric geometry consisting of
three atoms connecting electrodes in both the (100) and (111)
directions, see Fig.~\ref{fig:1}. The electrode-electrode distance is
in both cases chosen to be 9.6{\AA} and the wire atoms are initially
relaxed at zero bias. We choose this particular geometry since here
the middle atom (2) can roughly be considered to have only two bonds
namely to the left atom (1) and the right atom (3).

%%% Calculated Charge and the voltage drop
When the voltage is applied we observe a substantial redistribution of
the electronic charge. At 0V the wires have a net negative charge of
about 0.2 excess electrons. About half of this resides on atom 2. At
2V the excess electrons on atom 2 are almost halved. This charge
redistribution results in an asymmetric voltage drop where the main
drop is between the negative electrode and atom 2 (see Fig.~\ref{fig:1}).

\begin{figure}[t]
\epsfig{figure=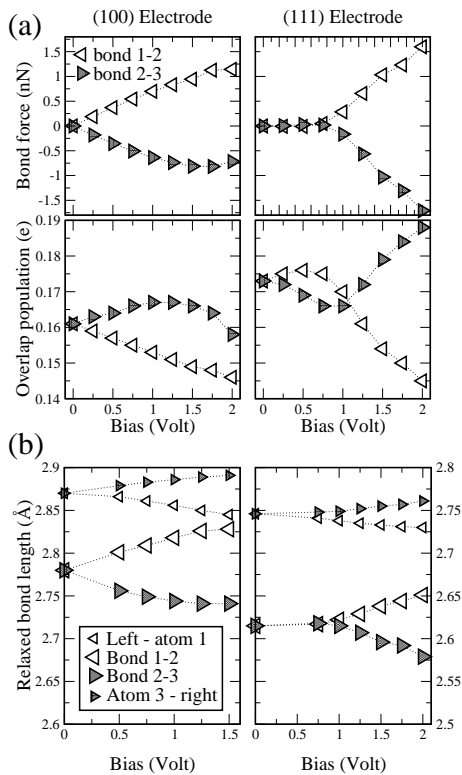,width=0.7\columnwidth, angle=0}
\caption{(a) The bondforce and overlap(bond) population for the (100) and
(111) connected wires. (b) The relaxed wire bondlengths for finite
bias.  }
\label{fig:2ab}
\end{figure}

%%%% Calculated force and bondlengths: FIG 2
In Fig.~\ref{fig:2ab}a we show the calculated bond forces for the
(100)/(111) wires with atomic positions fixed at the relaxed
values\cite{relax} obtained for zero bias. We find that the forces on
the symmetric atoms 1 and 3 roughly follow each other, also beyond the
linear bias regime\cite{ToHoSu00}. We note that the order of magnitude
and direction of the bond forces at 2V are similar for (100)/(111)
while the detailed behavior differs: the bond forces for the (100)
wire stay linear up to 1V and level off around 2V while we observe an
onset of the bond forces for (111) at 1V. The magnitude of the bond
forces of the order of 1-1.5nN for 1-2V is sizable compared to the
force required to break a one-atom thick wire $\sim 1.5$nN
\cite{RuBaAg01}.

Along with the bond force we also show the overlap population (OP) in
Fig.~\ref{fig:2ab}a. The OP is a measure of the electronic charge
residing in the bond. We note that except for the (111)-wire at low
bias, the force and OP curves are almost mirrored. The result that an
increase (decrease) in bond charge makes the bond force
compressive (repulsive) is what we expect intuitively. In
Fig.~\ref{fig:2ab}b we show the bond lengths after subsequent
relaxation of the wire atoms for finite bias. The main relaxations
take place in the wire bonds with smaller displacements in the 
wire-electrode bonds.

%%% General Theory of force
We aim at an qualitative understanding of the results for the
dependence of the force on the voltage bias. To this end we focus on
the OP. We can express\cite{forceprac} the force acting on atom $i$
due to the valence electrons using the force operator, ${\bf F}_{i}$, and the density
matrix ${\bf D}$,
\begin{equation}
\vec{F}_{i} = \mbox{Tr}[\vec{\bf F}_i \,{\bf D }] \,\,\, {\mbox{where}}, \,\,
\vec{\bf F}_{i} = -\frac{\partial{\bf H}}{\partial \vec R_i} \,.
\label{eq:genF}
\end{equation}
%CHEMICAL PICTURE OF BONDING AND COOP
Consider the bonding between two atoms represented by orbitals
$|\phi_1\rangle$ and $|\phi_2\rangle$ and separated by bond length
$b$. From Eq.~\ref{eq:genF} we get the bond force,
\begin{equation}
F_{\text{bond}} = -2 \left( 
\frac{\langle\phi_2 |H'|\phi_1\rangle}
{\langle\phi_2|\phi_1\rangle} \right)
\, O_{12}
\, , \,\,\,H'=\frac{\partial H}{\partial b}\,,
\end{equation}
where we assume that only the hopping element,
$\langle\phi_2|H|\phi_1\rangle$, changes with $b$. The bond force is
proportional to the OP for the 12 bond,
\begin{equation}
O_{12}=2\, {\bf S}_{12}\,{\bf D}_{12}\,\,.
\end{equation}
The OP is typically taken as a simple measure of the strength of a
chemical bond as suggested by Mulliken for molecules\cite{Mulliken55}.
For extended systems the density matrix can be expressed in terms of
the density of states (DOS) matrix, ${\brho}$. 
In equilibrium we have,
\begin{equation}
O_{12} = 2\,{\bf S}_{12}\,\int_{-\infty}^{\infty} d\varepsilon\,\, 
{\brho}_{12}(\varepsilon)\, n_F(\varepsilon-E_F)\,,
\label{eq:ovlEQ}
\end{equation}
where $n_F$ and $E_F$ is the Fermi function and energy.  The
contribution from states at different energy($\varepsilon$) in the
extended system to the bonding between the two orbitals is described
qualitatively by the OP weighted DOS (OPWDOS) or COOP curve, $2{\bf
S}_{12}\,{\brho}_{12}(\varepsilon)$, as discussed by
Hoffmann\cite{Hoffmann88}.

%NONEQUILIBRIUM
Now we consider the nonequilibrium situation where an electrical
current is running though a contact connecting two reservoirs, left
and right($L$,$R$), with different chemical potentials($\mu_L$,
$\mu_R$). The contact contains the 12 bond.  In this case the bond
force will change since the density and Hamiltonian matrices deviate
from equilibrium values.  The appropriate density matrix for the
nonequilibrium situation is constructed from scattering states, and
the total spectral density matrix(${\brho}$) is split into partial contributions
corresponding to their left(${\brho}^L$) or right(${\brho}^R$) origin (see
e.g. \cite{BrMoOr02}),
\begin{equation}
{\brho}(\varepsilon)=
{\brho}^L(\varepsilon)+{\brho}^R(\varepsilon)\,.
\label{eq:sumLR}
\end{equation}
The density matrix for nonequilibrium is then found by filling the
left and right originating states according to the respective chemical
potentials\cite{ToHoSu00,BrMoOr02},
\begin{equation}
{\bf D}= \int_{-\infty}^{\infty} d\varepsilon\, 
{\brho}^L(\varepsilon)n_F(\varepsilon-\mu_L)+ 
{\brho}^R(\varepsilon)n_F(\varepsilon-\mu_R) \,.
\label{eq:Dspec}
\end{equation}
We split the overlap population(Eq.~\ref{eq:ovlEQ}) further into its
partial left and right components,
\begin{equation}
O_{12}=O_{12}^L + O_{12}^R \,,
\end{equation} 
where
\begin{equation}
O_{12}^L = 2\,{\bf S}_{12}\,\int_{-\infty}^{\infty} d\varepsilon\,\, 
{\brho}_{12}^L(\varepsilon)\, n_F(\varepsilon-\mu_L)\,,
\label{eq:ovlNEQ12L}
\end{equation}
and likewise for $R$.  When more orbitals are involved on atom 1 and 2
the overlap populations can be broken down in orbital components. In
this case each orbital overlap will contribute with different weight
to the bond force according to their different force matrix elements
$H'$. This complication, together with the neglect of change in $H'$
with applied voltage, makes the bond force and OP only roughly
proportional, as we observe for the gold wires in Fig.~\ref{fig:2ab}.
However, as a first approximation, we can explain the observed change
in bond forces from the change in OP, which in turn means the change in
left ($\Delta O^L$) and right ($\Delta O^R$) contributions.

\begin{figure}[t]
\epsfig{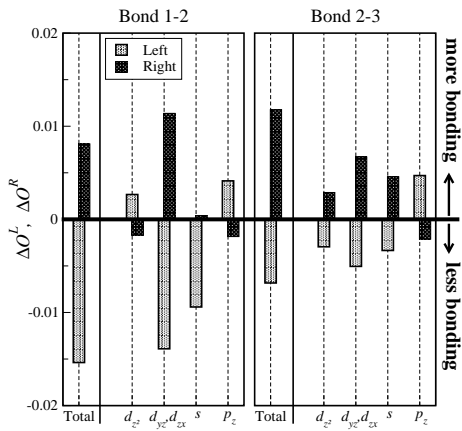}
\caption{The change in left and right partial OP, $\{ O_{12}^L,
O_{12}^R \}$ and $\{ O_{23}^L,O_{23}^R \}$ for a change in bias from
0V to 1V for the (100) wire. These are broken into main orbital 
contributions with respect to the middle atom (2). We show the sum of
the contributions for the degenerate $d_{yz}$ and $d_{zx}$.  }
\label{fig:3}
\end{figure}

In Fig.~\ref{fig:3} we consider the change from 0V to 1V in left and
right OP of the 12 and 23 bonds for the (100) case (we see a quite
similar pattern for (111)). For both bonds $\Delta O^L$ is negative
(decreasing bonding) while $\Delta O^R$ is positive (increasing
bonding). We have resolved these further into the orbitals on atom 2
and display the main contributions in Fig.~\ref{fig:3}. 

\begin{figure}[tb]
\epsfig{figure=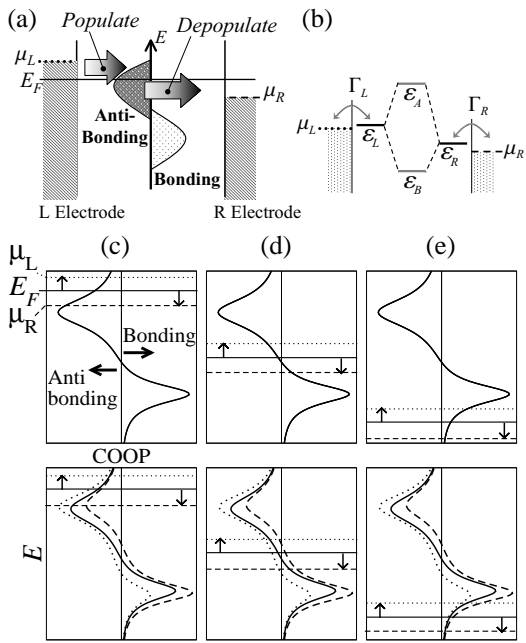,width=0.8\columnwidth}
\caption{
(a) Generic picture of the COOP(x-axis) curve and chemical
potentials of the electrodes, $\mu_L,\mu_R$, deviating from the
equilibrium, $\mu_L=\mu_R=E_F$.
(b) Simple model with a single bond between two atomic
orbitals($\varepsilon_L$, $\varepsilon_R$) forming
bonding($\varepsilon_B$) and anti-bonding states($\varepsilon_A$).
Atom $L$ couples to the left electrode by $\Gamma_L$(life-time
$\hbar/\Gamma_L$) and likewise for atom $R$.  
(c-e) COOP curves for different initial filling factors.  
Left(Right) quantities are denoted by dotted(dashed) lines; 
zero voltage quantities are solid.  The top(bottom) panels represent
zero(full) voltage drop between the atoms,
$\varepsilon_L=\varepsilon_R$($\varepsilon_L-\varepsilon_R=\mu_L-\mu_R$).}
\label{fig:4}
\end{figure}

%%%%%%%%%%%%%%%%%%% SIMPLE MODEL
In order to understand the physics involved we present in
Fig.~\ref{fig:4}a a simple picture. In the presence of current the
electrons from the left electrode populate states above the original
Fermi energy, $E_F$. These populate bonding or anti-bonding states
depending on the position of $E_F$. Likewise there is an depopulation
in electrons from the right. To make this picture more explicit we
consider in Fig.~\ref{fig:4}b a single bond between two atoms each
with a single atomic level, $\varepsilon_L$ and $\varepsilon_R$. The
levels couple via a matrix element $t<0$, and an overlap $S>0$, and
form bonding/anti-bonding orbitals, $\varepsilon_B$ and
$\varepsilon_A$. The left and right atoms are coupled to left and
right electrodes modelled by a level broadening $\Gamma_L$ 
of the left level and $\Gamma_R$ for the right. 
For this bond the resulting partial DOS due to the states
originating from the left is,
\begin{widetext}
\begin{equation}
\rho^L(\varepsilon)=
\frac{2}{\pi}
\frac{\Gamma_L\,\left(\varepsilon_R - \varepsilon \right) ( \varepsilon\,S + |t|)
}
{
\left[ 
(1 - S^2)( \varepsilon - \varepsilon_A) \,(\varepsilon - \varepsilon_B)
-\Gamma_L\Gamma_R
\right]^2 
+ 
\left[
\Gamma_R\left( \varepsilon - \varepsilon_L \right)  + 
\Gamma_L\left( \varepsilon - \varepsilon_R \right) 
\right]^2 
} \,\,.
\label{eq:rho2L}
\end{equation}
\end{widetext}
We note: (i) $\rho^R$ is found by exchanging $L\leftrightarrow R$.
(ii) $\rho^L$ is proportional to the coupling $\Gamma_L$. (iii) The
cross-over between bonding(positive) and anti-bonding(negative) is
given by the position of $\varepsilon_R$.  From (i), (iii) we see that
$\rho_L$ contains more anti-bonding character than $\rho_R$ when
$\varepsilon_L > \varepsilon_R$.

%%SIMPLE MODEL: NO VOLT DROP
We have plotted $L$/$R$ COOP curves in Fig.~\ref{fig:4}(c-e) for
different fillings and $\Gamma_L=\Gamma_R$. 
In the top row we assume that there is no voltage
drop between the atoms, $\varepsilon_L=\varepsilon_R$. Here the change
in OP comes entirely from the change in filling due to the
shift of $\mu_L$ and $\mu_R$ away from the equilibrium $E_F$
(Eq.~\ref{eq:ovlNEQ12L}). For almost filled states
(Fig.~\ref{fig:4}c, top) the lowering of $\mu_R$ deplete states
with anti-bonding character and $O^R$ increases. The increase in
$\mu_L$ leads to a decrease in $O^L$ which does not balance the
increase in $O^R$. The net effect is a {\em decrease} in anti-bonding
character and strengthening of the bond. When the states are almost
empty (Fig.~\ref{fig:4}e, top) we {\em also} observe a
strengthening since now $O^L$ increases more than $O^R$ is decreasing.
For a bond initially with maximal strength (Fig.~\ref{fig:4}d, top)
and, in general, for $E_F$ located in a region where the COOP
decreases, the bond weakens since both $O^R$ and $O^L$ decrease.

%%%SIMPLE MODEL:VOLT DROP 

In the bottom row of Fig.~\ref{fig:4}(c-e) we assume that the full
voltage drop takes place between the two atoms,
$\varepsilon_L-\varepsilon_R=\mu_L-\mu_R > 0$. The bonding peak becomes
dominating in the right COOP and likewise for the anti-bonding peak in
the left COOP. 
% We note that the total area under the left plus right
% COOP curves is constant, $2S^2/(S^2-1)$. 
The net result is less bond strengthening in (c) and (e), 
and more bond weakening in (d). The
largest effect of a voltage drop is seen for half filling
(Fig.~\ref{fig:4}d, bottom) where a large negative $\Delta O_L$ occurs because
the bonding peak decreases and at the same time more anti-bonding
states are being occupied. 

In summary, for equal coupling, $\Gamma_L =
\Gamma_R$, our simple model yield: (i) bond strengthening for almost
empty/filled states, (ii) bond weakening for about half filled states,
and (iii) a bond with a voltage drop is weaker than it would be
without a drop. When the bond is not coupled equally well to left and
right, $\Gamma_L \neq \Gamma_R$, the left and right overlap
contributions have to be weighted accordingly.

%% SIMPLE MODEL -> REAL WIRES
In the case of the gold wires we can to a good approximation assume
that the $d_{yz}$, $d_{zx}$, and $p_{z}$ orbital on atom 2 only couple
with the corresponding orbital on atom 1 and 3. The $d_{yz},d_{zx}$
states are almost filled and match the case in Fig.~\ref{fig:4}c:
$\Delta O^R>0$ while $\Delta O^L<0$. On the other hand the $p_z$ 
states are almost empty and match the situation in Fig.~\ref{fig:4}e:
$\Delta O^L>0$ while $\Delta O^R<0$. This is exactly the behavior we
observe in Fig.~\ref{fig:3}. For the $s$ orbitals, which are slightly
more than half-filled, we can expect a decrease in $O^L$, which will be
especially pronounced if a voltage drop takes place in the bond (cf.\
Fig~\ref{fig:4}d, bottom). This is also what we observe for the 12 bond in
Fig.~\ref{fig:3}. 

Based on our simple model we see that the almost completely filled
states($d$) and slightly more than half-filled states($s$) together
yield a {\em decrease} in $O^L$ and {\em increase} in $O^R$. Since
bond 12 couple more to the left electrode the total change in $O^L$
will be bigger than the change in $O^R$. This explains the bond
weakening of bond 12. The opposite is the case for bond 23. Also the
voltage drop taking place in bond 12 will weaken this bond with
respect to bond 23.

%%%%%%%%%%%%%%%% CONCLUSION

In conclusion we have analyzed first principles calculations of the
current-induced forces in an atomic gold wire. We have shown that the
current-induced forces are due to a shift in the population of bonding
and antibonding levels, and can be explained by 
examining the change in electrode decomposed bond charges(overlap
population) with applied voltage. Although we have concentrated on a
simple system the approach should apply to atomic scale conductors in
general.

%%%%%%%%%% acknowledgements
\hspace{0.25cm}
\begin{acknowledgments} 
We acknowledge support from the Danish Research Councils (M.B. and
K.S.), European Union (SATURN project) and MCyT (J.L.M. and P.O).  We
also thank the $\Psi_k$ network and CEPBA for funding collaborative
visits.
\end{acknowledgments} 
\hspace{0.25cm}
\small
%\bibliographystyle{revtex}
%\bibliography{refforce}

\begin{thebibliography}{18}
\bibitem[{met()}]{method}
\bibinfo{note}{We employ the {\sc Siesta} package\cite{SoArGa02} extended to
  the nonequilibrium situation by the {\sc TranSiesta} method\cite{BrMoOr02}.}
\bibitem[{rel()}]{relax}
\bibinfo{note}{The force is less than
  $0.02$~nN on all wire atoms.}
\end{thebibliography}

\end{document}